\newif\ifsingle

\ifsingle
\documentclass[11pt,draftclsnofoot, onecolumn]{IEEEtran}		
\else		
\documentclass[10pt,final, twocolumn]{IEEEtran}
\fi

\usepackage{times}
\usepackage{amsmath,dsfont}
\usepackage{amssymb,amsthm}
\usepackage{epsfig,verbatim}
\usepackage{subfigure}
\usepackage{setspace}
\usepackage{color}
\usepackage{cite}
\usepackage{epstopdf}
\usepackage{graphics}
\usepackage{accents}
\usepackage{acronym}
\usepackage[bookmarks,colorlinks]{hyperref}
\usepackage{booktabs}
\usepackage{mathtools}
\usepackage{algorithm}
\usepackage{algorithmic}
\usepackage{enumitem}
\usepackage{pstool}
\usepackage[linewidth=0.5pt]{mdframed}

\ifsingle

\newcommand{\figSpace}{\vspace{-0.2cm}}

\else

\newcommand{\figSpace}{\vspace{-0.3cm}}
\fi 

\acrodef{adc}[ADC]{Analog-to-Digital Convertor}
\acrodef{dac}[DAC]{digital-to-analog convertor}
\acrodef{cs}[CS]{Compressed Sensing}
\acrodef{dtft}[DTFT]{discrete-time Fourier transform}
\acrodef{dnn}[DNN]{deep neural network} 
\acrodef{csi}[CSI]{channel state information}
\acrodef{map}[MAP]{maximum a-posteriori probability}
\acrodef{snr}[SNR]{signal-to-noise ratio}
\acrodef{sinr}[SINR]{signal-to-interference-and-noise ratio}
\acrodef{bs}[BS]{Base Station} 
\acrodef{em}[EM]{electromagnetic} 
\acrodef{iot}[IOT]{Interent of Things}
\acrodef{mimo}[MIMO]{multiple-input multiple-output}
\acrodef{mse}[MSE]{mean-squared error}
\acrodef{pdf}[PDF]{probability density function}
\acrodef{rv}[RV]{random variable}
\acrodef{fec}[FEC]{forward error correction}
\acrodef{rs}[RS]{Reed-Solomon}
\acrodef{lti}[LTI]{linear time-invariant}
\acrodef{wss}[WSS]{wide-sense stationary}
\acrodef{psd}[PSD]{power spectral density}
\acrodef{ser}[SER]{symbol error rate} 
\acrodef{ber}[BER]{bit error rate} 
\acrodef{isi}[ISI]{intersymbol interference}  
\acrodef{awgn}[AWGN]{additive white Gaussian noise} 
\acrodef{ut}[UTs]{User Terminals} 
\acrodef{mmw}[mmWave]{millimeter wave}
\acrodef{ris}[RIS]{reconfigurable intelligent surface} 
\acrodef{dma}[DMA]{Dynamic Metasurface Antenna} 
\acrodef{5G}{fifth generation}
\acrodef{pa}[PA]{power amplifier}

\IEEEoverridecommandlockouts

\title{6G Wireless Communications: From Far-field  \\  Beam Steering to Near-field Beam Focusing}

\author{  
	\IEEEauthorblockN{Haiyang Zhang,~\IEEEmembership{Member,~IEEE}, Nir Shlezinger,~\IEEEmembership{Member,~IEEE}, Francesco Guidi,~\IEEEmembership{Member,~IEEE},\\ Davide Dardari,~\IEEEmembership{Senior Member,~IEEE}, and Yonina C. Eldar,~\IEEEmembership{Fellow,~IEEE}\\
	} 

}
 
\begin{document}
	
	\maketitle
 	\pagestyle{plain}  
\thispagestyle{plain} 


\begin{abstract}
 6G networks will be required to support higher data rates, improved energy efficiency, lower latency, and more diverse users compared with 5G systems. To meet these requirements, electrically extremely large-scale antenna arrays
 are envisioned to be key physical-layer technologies. 
As a consequence, it is expected that some portion of future 6G wireless communications  may  take place  in the radiating near-field (Fresnel) region, in addition to the far-field operation as in current wireless technologies.
 In this article, we discuss the opportunities and challenges that arise in radiating near-field communications. We begin by discussing the key physical characteristics of near-field communications, where the standard plane-wave propagation assumption no longer holds, and clarifying its implication on the modelling of wireless channels. Then, we elaborate on the ability to leverage spherical wavefronts via  beam focusing, highlighting its advantages for 6G systems. We point out several appealing  application scenarios which, with proper design, can  benefit from near-field operation, including interference mitigation in multi-user communications, accurate localization and focused sensing, as well as  wireless power transfer with minimal energy pollution. We conclude by discussing some of the design challenges and research directions that are yet to be explored to fully harness the potential of near-field operation.
\end{abstract}

\section{Introduction}

Along with the commercial deployment of  fifth-generation (5G) networks, the next sixth-generation (6G) wireless networks are gradually evolving from a vision into concrete designs. Notably, 6G networks will be required to: (i) support immense throughput, with peak data rates on the order of terabits per
second; (ii) support ultra-massive communications, with connection density of over 1 million per square kilometre; (iii) be highly energy efficient, with some applications expected to provide at least 1 terabit per Joule; (iv) and operate with ultra low latency, which in systems such as industrial control should be as low as 1 microsecond  \cite{rajatheva2020white}.
These performance requirements are necessary to support exciting new technologies of the 6G era, including the internet of everything, autonomous vehicles, and tele-medicine, in addition to conventional high-rate wireless communications.

\begin{figure*}
		\centering
		\includegraphics[width= 0.7\linewidth]{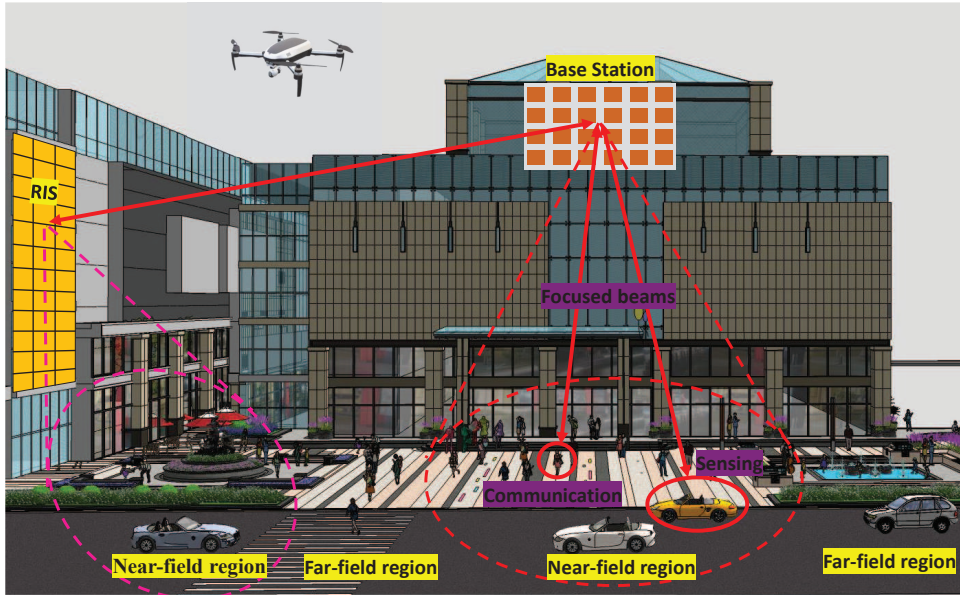}
		\figSpace
		\caption{An illustration of a near-field wireless communication system. The radiating near-field region can be up to tens of meters when the system operates in the mmWave frequencies with a large antenna array or \ac{ris}. }
\label{fig:model}
\end{figure*}

In order to support the ambitious requirements of 6G wireless networks, extremely large-scale (XL) antenna arrays, deployed with hundreds or even thousands of antennas at \acp{bs} or at passive \acp{ris}, are expected to be used. Also,  high-frequency spectra, i.e., millimeter wave (mmWave) and sub-terahertz (THz) bands, can provide large available bandwidth to support higher data rates, even though they might be affected by the limited coverage range. 
Consequently, future 6G networks are envisioned to utilize high-frequency bands, in addition to traditional sub-six GHz bands.
 The deployment of extremely large antenna arrays, especially in high-frequency bands, implies that future 6G wireless communications may take place in the near-field region in addition to the far-field, as illustrated in Fig. \ref{fig:model}. The boundary between the  near-field region and the far-field region is dictated by the Fraunhofer distance
   (also called Rayleigh distance) \cite{nepa2017near}.  The Fraunhofer distance represents the minimum distance for guaranteeing the phase difference of received signals across the array elements of at most 22.5 degrees~\cite{Haiquan}, which is considered the limit under which wave propagation can be viewed as planar. The Fraunhofer distance is proportional to the square of the antenna aperture and inversely proportional to the signal wavelength. As the antenna aperture of 6G transmitters increases,  the boundary can be  up to several dozen of meters. For example, for a  planar array  with diameter of 0.5 meters
transmitting at a carrier frequency of 28~GHz, the radiating near-field distance is up to 47~meters.
In addition, the near-field region can be further divided into radiating near-field region and reactive near-field region. As the reactive near-field region corresponds to signaling in extremely close ranges (in the order of the wavelength), in this paper we henceforth indicate with near-field the radiating near-field, being the relevant one for wireless communications. 

While conventional wireless communications operate in the far-field, where \ac{em} wavefronts can be reliably approximated as planar, in the near-field region the spherical shaping of the wavefront cannot be neglected. This property brings forth new opportunities and challenges for future wireless communications systems design. Spherical wavefronts can be exploited to generate focused beams in specific spatial region, namely {\em beam focusing},
which is not achievable with traditional far-field beam steering, where signals can only be pointed towards a specific direction.
While beam focusing can be leveraged to facilitate multi-user communication by, e.g., enabling new levels of interference mitigation \cite{zhang2021beam, lu2022near}, it also gives rise to new design and signal processing challenges. This stems from the fact that the near-field spherical wavefront is inherently different from that observed in conventional far-field designs, and thus existing wireless communications models, schemes and results derived assuming far-field operation may no longer be valid. 
This indicates the need  to study  the properties, potential benefits, and new design challenges, which arise from the radiating near-field operation of  6G wireless communications. 


In this article, we provide a general overview of the opportunities and challenges for future 6G systems likely operating in the near-field region. We first present the distinct physical features of radiating near-field wireless communications, including spherical wavefronts and distance-aware channel models. 
We then detailed
the principle
of exploiting the spherical wavefronts of the signals to implement {\em beam focusing}, 
and identify its potential advantages 
over conventional
far-field beam steering. To show the potential benefits of beam focusing for 6G systems, we discuss three typical applications which rely on EM signaling and possibly operate in the near-field region: (i)  multi-user communications; (ii)  localization and sensing; and (iii)  wireless power transfer. In addition, using the multi-user communication operated in the near-field as an example, we numerically demonstrate the interference mitigation ability of beam focusing, showing that a near-field aware design allows exploiting  new degree of freedom (DoF) to mitigate co-channel interference in both angle and distance domains.

Next, we highlight key design challenges in near-field communications and discuss the corresponding potential solutions and research directions. We concentrate on several unexplored algorithmic challenges imposed by the requirement to implement beam focusing for 6G networks, including  near-field channel estimation; mis-focusing/beam split effect in near-field wideband communications.  In addition, we discuss the opportunities and challenges of exploiting the high-rank property of near-field line-of-sight  (LOS) \ac{mimo} channels to enhance the MIMO multiplexing gain, as well as hardware implementation challenges.



\section{Radiating Near-field: Physical Features}
\subsection{Spherical Versus Planar Waves}
\label{subsec:spherical}
\begin{figure} 
\centering 
 \includegraphics[width=3.4in]{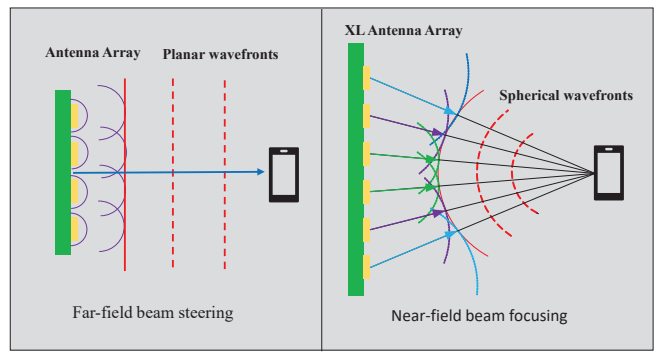} 
\caption{Illustration of the principles of far-field beam steering and near-field beam focusing.}
   \label{fig:principle}
\end{figure}

As
widely known, when antennas radiate \ac{em} waves in the surrounding free-space, they propagate exhibiting a spherical wavefront. However, in traditional wireless communications, the wavefront can be well approximated as being planar due to the large distances involved with respect to the operating wavelength. Nonetheless, such an approximation no longer holds in near-field radiative conditions
and, consequently, the wavefront impinging on the  receiver is spherical.
This peculiarity, visualized in Fig.~\ref{fig:principle}, indicates that one can associate more information to an {EM} wave compared with far-field planar waves, bringing forth the possibility to improve communication performance or to enable other applications relying on EM radiation thanks to the availability of more diverse radiation patterns.
Indeed,  incident spherical wavefronts carry not only angular information, as typically happens in the far-field regime, but also distance information. This property can be exploited to, e.g., determine the position of a transmitting source without requiring an ad-hoc synchronization procedure and the subsequent exchange of several messages, as demonstrated in \cite{guidi2019radio}. This position-awareness affects the channel model for near-field wireless communications, as discussed next.



\subsection{Near-Field Channel Model}
\label{subsec:channel}

The near-field channel model has been mainly considered and analyzed for free-space propagation.
 In classic far-field free-space channel models, the propagation distances from each antenna element to one target user are approximately identical, yielding the same path gain. Furthermore, antenna arrays share identical angle of arrival/departure and, thus, the phase of each element in the array steering vector can be modelled as linear to the antenna index \cite{cui2021channel}.   Unlike far-field free-space channel models,  in the near-field the aforementioned approximations do not hold, and  the antenna elements have different distances with the target user and different angles of arrival/departure. Thus,  in the near-field channel model, the wireless link from each antenna element to the target user has different path gain and phase variations \cite{Haiquan}.


The characterization of multipath environments gives rise to additional channel modelling challenges.
As previously discussed, 6G fosters the adoption of electrically large antenna arrays with hundreds or even thousands of radiating elements.
Thus, in contrast to traditional communication systems,  
the large array aperture  makes  the  spatial channel properties experienced across the  array  non-stationary \cite{pizzo2021spatial}. This means that some multipath components might be detected by certain elements of the array but not from the others and, thus, each set of antennas requires proper channel characterization. Traditional fading models are no longer suitable for describing such non-stationary channels.
In addition, some of the surrounding scatterers may reside in the radiating near-field with respect to either the transmitter or the receiver, while others might lie in the far-field, so that hybrid models should be accounted for while representing the channel characteristics. 
Recently proposed modelling approaches, relying on  Fourier plane-wave series expansion of the channel response  \cite{pizzo2021spatial} or by representing the environment using discrete coupled dipoles \cite{faqiri2022physfad}, allow to implicitly account for such hybrid scattering scenarios. Nonetheless, complete modelling of near-field wireless channels is still an active area of research.

\section{Beam focusing}
\subsection{From Beam Steering to Beam Focusing}

In conventional far-field  wireless communications,
transmit {\em beam steering} refers to an array signal processing technique where a multi-antenna transmitter sends EM signals to a specific direction, as shown in the left side of Fig.~\ref{fig:principle}.
Transmit beam steering is a key method for improving spectrum efficiency and controlling co-channel interference in far-field communications. However, as the EM wavefronts are planar, the transmitter can only control the relative angle towards which most of the energy is radiated.

Differently, in the radiating near-field region, {\em beam focusing} takes advantage of the non-negligible spherical wavefront to focus the radiated energy in a specific spatial location, i.e., not only by angle, but also by a specific depth along the direction of propagation, as shown in the right side of  Fig.~\ref{fig:principle}.  

Indeed, similarly to beam steering, beam focusing is a transmit technique based on precoding of the outgoing signal to achieve a desired radiation pattern. In particular, in order to properly generating focused beams, one must separately weight the spherical wave signals from each individual antenna such that they are added constructively at the desired focal point, yielding high signal strength, and have the radiated spherical wave signals added destructively in other points (e.g., to suppress interference). As in conventional beam steering, the most flexible design is achieved using fully-digital arrays, where one can individually control the outgoing wavefront at each antenna, while typically resulting in costly and power-hungry designs for large-scale antennas and high-frequency radiation. Nonetheless, beam focusing can also be achieved using hybrid analog/digital techniques and even with purely analog beam steering using, e.g., phased arrays, though typically with reduced flexibility compared with fully-digital arrays, especially when aiming for complex patterns with multiple focal points \cite{zhang2021beam}.

The ability to generate focused beams can be exploited to facilitate future 6G wireless communications. The advantages of beam focusing when working in the near-field include: 
\begin{itemize}
\item {\em Beam focusing provides a new DoF to  shape the  interference.} Beam focusing can not only control multi-user interference in the angle domain, as traditional beam steering, but also control the interference in the distance domain.
\item{\em Beam focusing enables capacity-approaching near-field MIMO communications}.  Capacity-approaching MIMO communication in near field leads to complex array phase profiles. As shown in \cite{decarli2021communication}, these can be efficiently approximated using a combination of simpler multiple focusing beams. 
\end{itemize}
In the next section we illustrate some specific application examples of these advantages. 

\subsection{Application Scenarios}
In the following, we list some of the typical scenarios in 6G networks where near-field aware beam focusing can be exploited to enhance performance:
\begin{itemize}
\item {\em Near-Field Multi-User Communications}: 
6G \acp{bs} and access points will support a multitude of users. Such application scenarios result in multiple spectrum sharing users, which in turn is likely to result in notable interference. When the users are located in the radiating near-field region of the \ac{bs}, co-channel interference can be mitigated via beam focusing, even when the users lie in the same relative direction \cite{zhang2021beam}. This interference mitigation ability, which is not present in conventional far-field communications, gives rise to the possibility to support multiple coexisting orthogonal links, even at the same angles,  thus allowing spatial user densification.
\item {\em Near-Field Localization and Sensing}: 
The diverse heterogeneous nature of 6G networks indicates that some devices will utilize EM radiation not solely for communications, but also for probing the environment, e.g., for sensing and/or RF localization.   Wireless localization and sensing are expected to be enhanced by the consideration of distance-aware channels. As an example, preliminary results have shown the possibility to perform holographic localization by exploiting the DoF offered by the spherical wavefront, so that positioning is boosted at an unprecedented scale thanks to the position information encapsulated in the wavefront and without the need to deploy more ad hoc nodes to triangulate signals \cite{guerra2021near}.
\item {\em Near-Field Wireless Power Transfer}:
Wireless Power Transfer (WPT) is an emerging technology that allows an energy transmitter to charge multiple remote devices wirelessly, and has many potential applications in 6G networks \cite{zhang2021near}. Two core challenges in WPT are associated with the fact that  energy transfer efficiency is relatively low due to the path loss degradation, and the need to avoid energy pollution, i.e., radiating energy at specific sensitive location. When WPT is carried out in the radiating near-field, beam focusing enables to jointly tackle these challenges,  allowing to achieve efficient power transfer with  minimal  energy pollution, as shown in \cite{zhang2021near}.
\end{itemize}

While the above applications arise from the main expected usages of EM radiation in 6G networks,  there are also many other exciting  application-oriented opportunities related to near-field communications.
First, some emerging technologies aim at combining EM radiation for dual purposes. These include integrated sensing and communications, where beam focusing can greatly contribute to facilitating co-existence, and simultaneous wireless information and power transfer. 
An additional communication aspect which can benefit from beam focusing is physical-layer security, which aims at exploiting the physical features of wireless channels to convey information while keeping it concealed from eavesdroppers. In the conventional far-field case,  beam steering and noise jamming are two commonly used signal processing tools to enhance the achievable secrecy rate. As the communication devices move to the radiating near-field, one can apply focused beams to focus the confidential messages on legitimate users and avoid information leakage to eavesdroppers. In this case, beam focusing can increase the secrecy rate and potentially enhance the energy efficiency of the whole system because jamming signals may be unnecessary.

\begin{figure} 
  \centering 
 \includegraphics[width=3.6in]{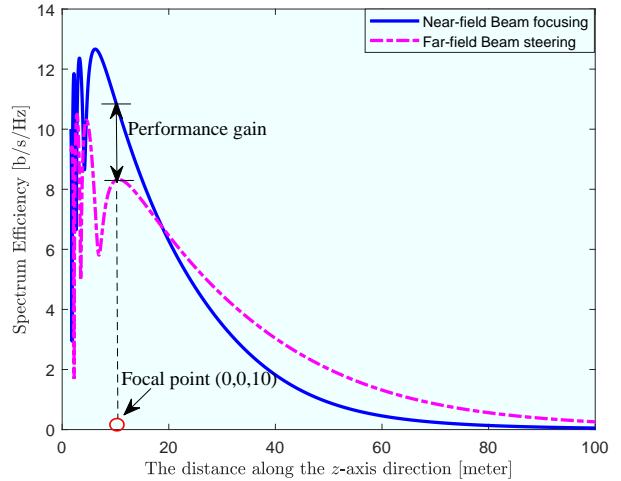} 
\caption{Comparison of beam focusing and beam steering in terms of spectrum efficiency.}
   \label{fig:rate_user1}
\end{figure}

\begin{figure} 
  \centering    \includegraphics[width=3.6in]{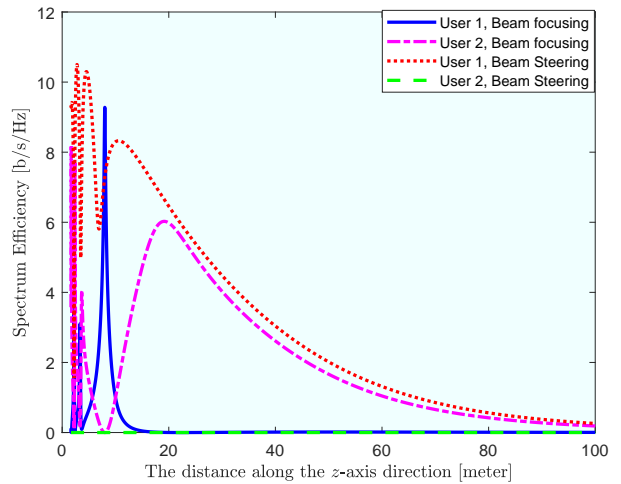} 
\caption{Spectrum efficiencies per user versus location along the  $z$-axis.}
  \label{fig:rate_twoUser} 
\end{figure}

\begin{figure*}
\centering
\includegraphics[width= 1\linewidth]{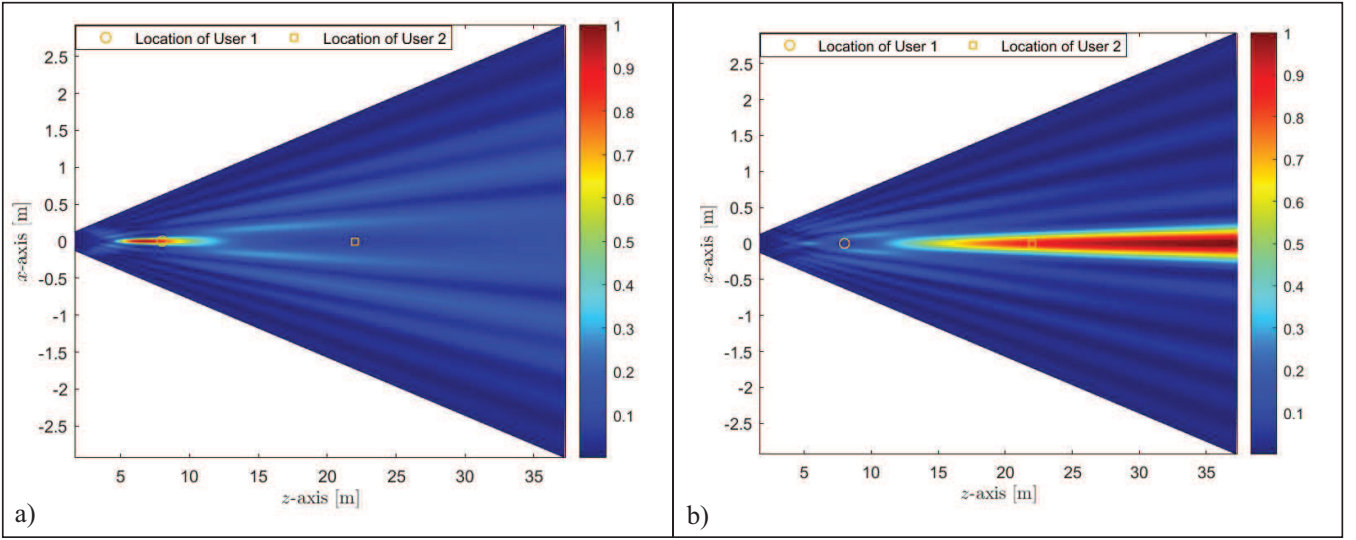}
\caption{The normalized signal power measurement of two focused beams: a) focused beam for user 1; b) focused beam for user 2. }
\label{fig:Beam_pattern}
\end{figure*}

\section{Numerical Results}
\label{ssec:Sim}
To demonstrate the potential of near-field beam focusing for 6G wireless communications, we simulate a multi-user communications setup, and present numerical results to illustrate the interference mitigation ability of beam focusing.  We consider a near-field multi-user communication scenario where the \ac{bs} deploys a fully-digital uniform planar array (UPA) positioned in the  $xy$-plane, and two single-antenna users are placed in the  $z$-axis, i.e., in the same angular direction. In practice, the antenna architecture imposes some limitations on the ability to generate different beam patterns, which obviously affect the beam-focusing capabilities of the transmitter. Here we consider the ideal case of fully digital architecture 
to study the performance limits of near-field-aware beam focusing, although constrained and hybrid antenna architectures may be more commonly used in future 6G networks. We consider the length and width of the UPA are 0.8 and 0.4 meters, the space between each element is one-wavelength, and the  carrier frequency is  $28$~GHz. 
Under such parameter
settings, the Fraunhofer distance, i.e., the limit of the radiating near-field region, is approximately  $149$ meters. This setting is used to exemplify  the beam-focusing abilities for near-field communications and does not target a specific peak data rate.  In our simulation, as in \cite{zhang2021beam, lu2022near}, we adopt the free-space line-of-sight channel model. The maximum transmit power is set to 10 dBm, noise power spectral density is -174 dBm per Hertz, and signal bandwidth is 100 MHz.  


We first illustrate the performance gain of beam focusing compared to beam steering for a single-user scenario.  Fig. \ref{fig:rate_user1} shows the spectrum efficiency when the receiver is located at different points along the   $z$-axis, achieved by the fixed beam focusing and beam steering antenna configurations. The transmit pattern used for beam focusing is based on the scheme proposed in \cite{zhang2021beam}, and is pointed at the focal point of  $(0,0,10)$~meters, located in the near-field region. Beam steering  is designed using classical antenna array design tools valid for far-field conditions to steer the beam towards the same direction, i.e., along the  $z$-axis, with the objective of maximizing the received power.  In Fig.~\ref{fig:rate_user1}, it is clearly shown that beam focusing can significantly increase the spectrum efficiency compared to beam steering when the user is located in the proximity of the focal point. This is because beam steering is designed for the far-field scenario, resulting in performance loss when operating in near-field users due to channel mismatch. Moreover, we can observe that when the user is located on the   $z$-axis but far away from the focal point, the spectrum efficiency of beam focusing is lower than that of far-field beam steering, which implies the ability of beam focusing on reducing radiating interference.

Next, we demonstrate the advantage of beam focusing in distinguishing different users with the same angular direction. 
As in \cite{zhang2021beam}, the focused beams are designed to maximize the sum spectrum efficiency of two near-field focal points located at coordinates (0,0,8) meters and  (0,0,22) meters. 
We also design the far-field steered beams towards the same direction. 
Fig.~\ref{fig:rate_twoUser} illustrates the spectrum efficiencies of each of the two users along the  $z$-axis (when one user is moving along the  $z$-axis, the other is fixed at the focal point), obtained using near-field beam focusing and far-field beam-steering antenna configurations.  We observe in Fig.~\ref{fig:rate_twoUser} that the peak spectrum efficiencies of each of the two users occur when they are located around their corresponding focal points, implying that the designed focused beams are all capable of yielding reliable communications with minimal degradation due to interference. In contrast, only one user can achieve a favorable spectrum efficiency for the far-field steer beams whereas the spectrum efficiency of the second user is approximately zero. This is because the two focal points have the same angular direction, and will generate strong interference, and thus  beam steering allocates all the transmit power to one user with the better channel (i.e., smaller path loss) to maximize the sum spectrum efficiency.

In Fig. \ref{fig:Beam_pattern} we visualize the co-channel interference mitigation ability of beam focusing. In particular,  we  plot the normalized signal power measurement of two focused beams on the  $xz$-plane, where the focused beams are designed and used as in Fig. \ref{fig:rate_twoUser}. 
From Fig. \ref{fig:Beam_pattern}, we can observe that  near-field focusing can not only enhance the signal strength at the focusing point, but also eliminate the co-channel interference to other users. For example, as shown in Fig. \ref{fig:Beam_pattern}(a), the energy of the beam designed for user 1 is mainly focused around user 1, and it generates negligible co-channel interference to user 2, even if the two users lies in the same angular direction.  Figs. \ref{fig:rate_twoUser} and \ref{fig:Beam_pattern} indicate that, by properly accounting for the expected near-field operation via dedicated beam focusing, it is possible to give rise to new levels of interference mitigation not achievable in the conventional far-field, as one can not only control the multi-user interference in the angle domain, but also balance interference in the distance domain.
This aspect is particularly appealing in realizing ultra-massive networks.

\section{Design challenges and research directions}
\label{sec:directions}

Beam focusing enables sending signals on target regions with weak signal power leakage on other regions, as stated above,  which makes it an appealing technique for various future 6G applications. Approaching the near-field beam focusing performance boost, however, necessitates carefully addressing several challenges, opening up numerous fascinating research opportunities and directions. Some of them are discussed  briefly below.


\subsection{Channel Estimation} 

The ability to achieve focused beams heavily relies on accurate knowledge of the wireless channel.
Since the aim is to have the signal energy concentrated on a small region around the desired focal points, beam focusing is more sensitive to  inaccurate channel knowledge than far-field beam steering. 
Consequently, a relevant research direction involves studying the accurate near-field channel estimation techniques, and designing robust beam focusing approach to cope with the CSI inaccuracy problem.



For far-field high-frequency communication systems such as 5G mmWave, the steering vector of the antenna array is only related to the angle of arrival, and thus wireless channel can be represented by a Fourier dictionary-based sparse model. However, for the near-field case, the array steering vector depends on both the angles of arrival and the user distance. Therefore, the Fourier dictionary which only samples in the angular-domain is not appropriate to model near-field channels. 
One elegant solution is the polar-domain near-field channel modeling scheme \cite{cui2021channel}, which has strictly proved the non-uniform
sampling criterion in the distance dimension for the first time.
Consequently, new channel estimation methods based on the polar-domain near-field channel modeling  are highly desired and an important research direction.





\subsection{Beam Misfocus/Split Issue}

In order to focus beams on the target locations in near-field communications, the phase of each antenna should be designed to compensate for the signal transmission delay from the element to the target focal point. As a result, the phase of each transmit antenna element should depend on the transmission distance and frequency/wavelength, i.e., one should use frequency-selective precoding to achieve focused wideband beams. However, in common phased array-based communication systems, phase shifters are usually designed to be approximately frequency-flat within the communication band. Attempting to realize near-field wideband beam focusing using frequency-flat phase shifter hardware results in the so-called beam misfocus  or beam split \cite{myers2020infocus} phenomenon, where the generated beams at different frequencies will focus on different locations. The misfocus effect limits the effective bandwidth of the phased array, resulting in poor performance of wideband communication systems. As a result, it is highly desirable to reduce or eliminate the beam misfocus problems, either via signal processing techniques or by design of flexible precoding hardware, to overcome the harmful effects of this phenomenon in near-field wideband communications.


 \subsection{MIMO Multiplexing Gain}
\label{ssec:MIMO}

The additional DoF of the near-field distance-aware channel model compared with its far-field counterpart gives rise to potential \ac{mimo} multiplexing gains. Unlike conventional far-field LOS \ac{mimo} channels, which are represented by rank-one matrices, the near-field LOS \ac{mimo} channel matrix, depending on the geometric configuration, can potentially become full-rank thanks to the richer information carried by spherical waves propagation.     
This implies that near-field LOS \ac{mimo} channels provide increased spatial DoF, which can be translated into multiplexing gain \cite{decarli2021communication}. 
The exploitation of this channel property depends on the ability to address several design challenges. For once, the corresponding optimal precoding and decoding matrices are very sensitive to the geometric configuration of antennas normalized to the wavelength, thus making the \ac{csi} estimation process as well as hardware constraints more critical, especially at high frequency. As previously outlined, optimal precoding/decoding can be efficiently approximated by beam focusing operations, thus reducing the requirements on hardware. 


 \subsection{Hardware Implementation}
\label{ssec:hardware}
 

The realization of extremely large-scale antenna arrays especially operating at high-frequency bands, which naturally results in near-field operation whose potential is discussed in this article, is still subject to a multitude of implementation challenges. 
 In particular, the hardware technology which can realize extremely large-scale antenna arrays with high-frequency signals is an area of ongoing  research. For example, the efficiency of power amplifiers is typically frequency-dependent and tends to decrease as the carrier frequency increases. A possible solution is to design energy-efficiency high-frequency hardware devices based on advanced semiconductor technology such as scaled SiGe bipolar technology for supporting 6G wireless networks~\cite{tataria2022six}. For example, ADC/DAC in traditional transceiver architecture will
consume an excessive amount of power for a terabit-per-second data rate. Therefore, one important
research opportunity is how to explore new architectures, like those implementing high-order modulation in analog/RF domain directly to overcome this
challenge~\cite{heydari2021terahertz}.
Although these important hardware implementation challenges are not specific to near-field communications, they need to be carefully addressed for implementing efficient 6G near-field communication systems in practice.


  

\section{Conclusion}
To date, wireless communications are mainly studied and designed in the far-field region, where the  wavefronts can be well-approximated as planar. However, for future 6G networks, some portion of devices may operate in the radiating near-field region, where the conventional plane wave propagation assumption in far-field is no longer valid. In this article, we provided an overview of the opportunities and challenges of radiating near-field wireless communication systems. We first presented the key characteristics of near-field radiation where the spherical waveform propagation model holds, and clarified the new properties of the near-field channel model. Then, we discussed the emerging near-field beam focusing capability, highlighted its advantages for wireless communications,  and pointed out its several appealing 6G-related applications. We concluded with some of the design challenges and potential research directions.
It is acknowledged that while near-field wireless communications are promising for 6G, more research is still necessary to address the daunting challenges for more realistic evaluation and possible commercialization.

	\bibliographystyle{IEEEtran}
	\bibliography{IEEEabrv,refs}


\begin{IEEEbiographynophoto}{Haiyang Zhang}
 is an Assistant Professor in the School of Communication and Information Engineering, Nanjing University of Posts and Telecommunications, China.

\end{IEEEbiographynophoto}	
\vskip -2\baselineskip plus -1fil

\begin{IEEEbiographynophoto}{Nir Shlezinger}  is an Assistant Professor in the School of Electrical and Computer Engineering in Ben-Gurion University, Israel. 
\end{IEEEbiographynophoto}
\vskip -2\baselineskip plus -1fil

\begin{IEEEbiographynophoto}{Francesco Guidi} (francesco.guidi@ieiit.cnr.it) received his Ph.D. degree in electronics, telecommunications, and information
technologies from Ecole Polytechnique Paris-Tech, France (computer
science specialty) and from the University of Bologna,
Italy. He is currently a researcher with IEIIT-CNR, Italy.
\end{IEEEbiographynophoto}
\vskip -2\baselineskip plus -1fill

\begin{IEEEbiographynophoto}{Davide Dardari} (davide.dardari@unibo.it) is a full professor in the Department of Electrical, Electronic, and Information Engineering "Guglielmo Marconi" (DEI) at the University of Bologna, and affiliate at WiLAB-CNIT, Italy. 
\end{IEEEbiographynophoto}	
\vskip -2\baselineskip plus -1fill


\begin{IEEEbiographynophoto}{Yonina C. Eldar} (yonina.eldar@weizmann.ac.il)
is a Professor in the Department of Math and Computer Science, Weizmann Institute of Science, Israel, where she heads the center for Biomedical Engineering and Signal Processing. She is a member of the Israel Academy of Sciences and Humanities, an IEEE Fellow and a EURASIP Fellow.
\end{IEEEbiographynophoto}	
	
\end{document}